\documentclass[11pt,twoside]{article}

\usepackage{asp2006}
\usepackage{epsf}
\usepackage{lscape}

\usepackage{natbib}

\usepackage{graphicx}

\newcommand{\s}{s$^{-1}$}

\def\ltsima{$\; \buildrel < \over \sim \;$}
\def\simlt{\lower.5ex\hbox{\ltsima}}
\def\gtsima{$\; \buildrel > \over \sim \;$}
\def\simgt{\lower.5ex\hbox{\gtsima}}
\def\gsimeq
{\hbox{\raise0.5ex\hbox{$>\lower1.06ex\hbox{$\kern-1.07em{\sim}$}$}}}
\def\lsimeq
{\hbox{\raise0.5ex\hbox{$<\lower1.06ex\hbox{$\kern-1.07em{\sim}$}$}}}

\def\xmm{{\it XMM-Newton }}

\def\xmm{{\it XMM-Newton}}

\def\pg1404{PG~1404+226}

\def\mnras{MNRAS}
\def\apj{ApJ}

\def\apjl{ApJL}
\def\apjs{ApJS}
\def\aap{A\&A}
\def\araa{ARA\&A}
\def\nat{Nature}
\def\pasj{PASJ}

\begin{document}
\title{Molecular Clouds: X-ray mirrors of the Galactic nuclear activity}   
\author{G. Ponti\altaffilmark{1,2}, R. Terrier\altaffilmark{1}, A. Goldwurm\altaffilmark{3,1}, 
G. Belanger\altaffilmark{4} and G. Trap\altaffilmark{3,1}}
\altaffiltext{1}{APC Universit\'e Paris 7 Denis Diderot, 75205 Paris Cedex 13, France}  
\altaffiltext{2}{School of Physics and Astronomy, University of Southampton, Highfield, 
SO17 1BJ, UK}  
\altaffiltext{3}{Service d'Astrophysique/IRFU/DSM, CEA Saclay, 
91191, Gif-sur-Yvette Cedex, France}
\altaffiltext{4}{ESA/ESAC, PO Box 78, 28691 Villanueva de la Ca\~nada, Spain}

\begin{abstract} 
We present the result of a study of the X-ray emission from the Galactic 
Centre (GC) Molecular Clouds (MC), within 15 arcmin from Sgr A*. We 
use \xmm\ data spanning about 8 years.
We observe an apparent super-luminal motion of a light front illuminating 
a MC. This might be due to a source outside the MC (such 
as Sgr A* or a bright and long outburst of a X-ray binary), while it can not be 
due to low energy cosmic rays or a source located inside the cloud. 
We also observe a decrease of the X-ray emission from G0.11-0.11, 
behaviour similar to the one of Sgr B2. The line intensities, clouds 
dimensions, columns densities and positions with respect to Sgr A*, are 
consistent with being produced by the same Sgr A* flare. 
The required high luminosity (about 1.5$\times10^{39}$ erg s$^{-1}$) 
can hardly be produced by a binary system, while it is in agreement 
with a flare of Sgr A* fading about 100 years ago. 
\end{abstract}

\section{Introduction}   

In the central few hundreds parsecs of the Milky Way a high concentration of 
MC is present (Morris \& Serabyn 1996; Bally et al. 1987; 
Tsuboi et al. 1999). Sunyaev et al. (1993; 1998) first realised that these clumps 
of material can behave like mirrors of past bright X-ray events occurring in the 
GC. Thus, several authors have studied the X-ray bright 
MC (Koyama et al. 1996; Murakami et al. 2001) to constrain the past activity 
of the GC and, in particular, of Sgr A*, the counterpart of the supermassive 
black hole at the GC (Sch\"odel et al. 2002; Gillessen et al. 2009). On the other hand, 
cosmic ray irradiation can explain the high X-ray emission from these MC equally 
well (Valinia et al. 2000; Yusef-Zadeh et al. 2002; 2007; Dogiel et al. 2009; 
Bykov 2003). Here we study the MC emission during the 8 years \xmm\ monitoring 
of the 15 arcmin around Sgr A*. 

\section{Spectral analysis}

We first define the MC selecting the brightest regions in the CS maps (Tsuboi et 
al. 1999). We then extract and simultaneously fit the EPIC-pn and MOS spectra.
We observe that in each MC a narrow and neutral Fe K$\alpha$+$\beta$ line, 
with equivalent widths of the order of 0.7-1 keV, is required.
The MC power law emission is extremely flat with a spectral index between 
$\Gamma\sim0.8-1$, moreover most of the MC require a neutral Fe K 
absorption edge ($\tau\sim0.2-0.4$). These features are the signature of  
a reflection component from cold matter (Nandra \& George 1994; Ponti et al. 
2006; 2009; Bianchi et al. 2009a,b). 

\subsection{G0.11-0.11}

The upper left panel of Figure 1 shows the Fe K line light curve 
of the MC, G0.11-0.11. The Fe K emission is clearly variable with a 
strong decrease of the order of 50 \% within the 8 years of \xmm\ monitoring.
A similar decline is shown by the Sgr B2 MC (Inui et al. 2009; Terrier et 
al. 2009a,b). Assuming that G0.11-0.11 is located at its minimal distance 
from Sgr A* and that it is illuminated by it, we estimate a Sgr A* 
luminosity of L$\geq$$10^{39}$ erg s$^{-1}$, occurring more than 
75 years ago (Sunyaev \& Churazov 1998; Amo-Baladron et al. 2009). 
These values are surprisingly similar to the ones inferred from the study 
of Sgr B2 (Koyama et al. 1996). Moreover both light curves are in a 
decay phase, with similar variations (Inui et al. 2009; Terrier et al. 2010a,b). 
We thus assume that the flare luminosity illuminating Sgr B2 and G0.11-0.11 
is exactly the same and thus we constrain the position of G0.11-0.11. It has 
to be about 17 pc behind the plane of Sgr A*. The upper right panel of 
Fig. 1 shows that in this location G0.11-0.11 is illuminated by 
the same light front hitting Sgr B2. Although, this does not 
rule out other possibilities for the Fe K emission, it is a strong constraint in favour 
of the single flare hypothesis.
\begin{figure}
\includegraphics[scale = 0.3]{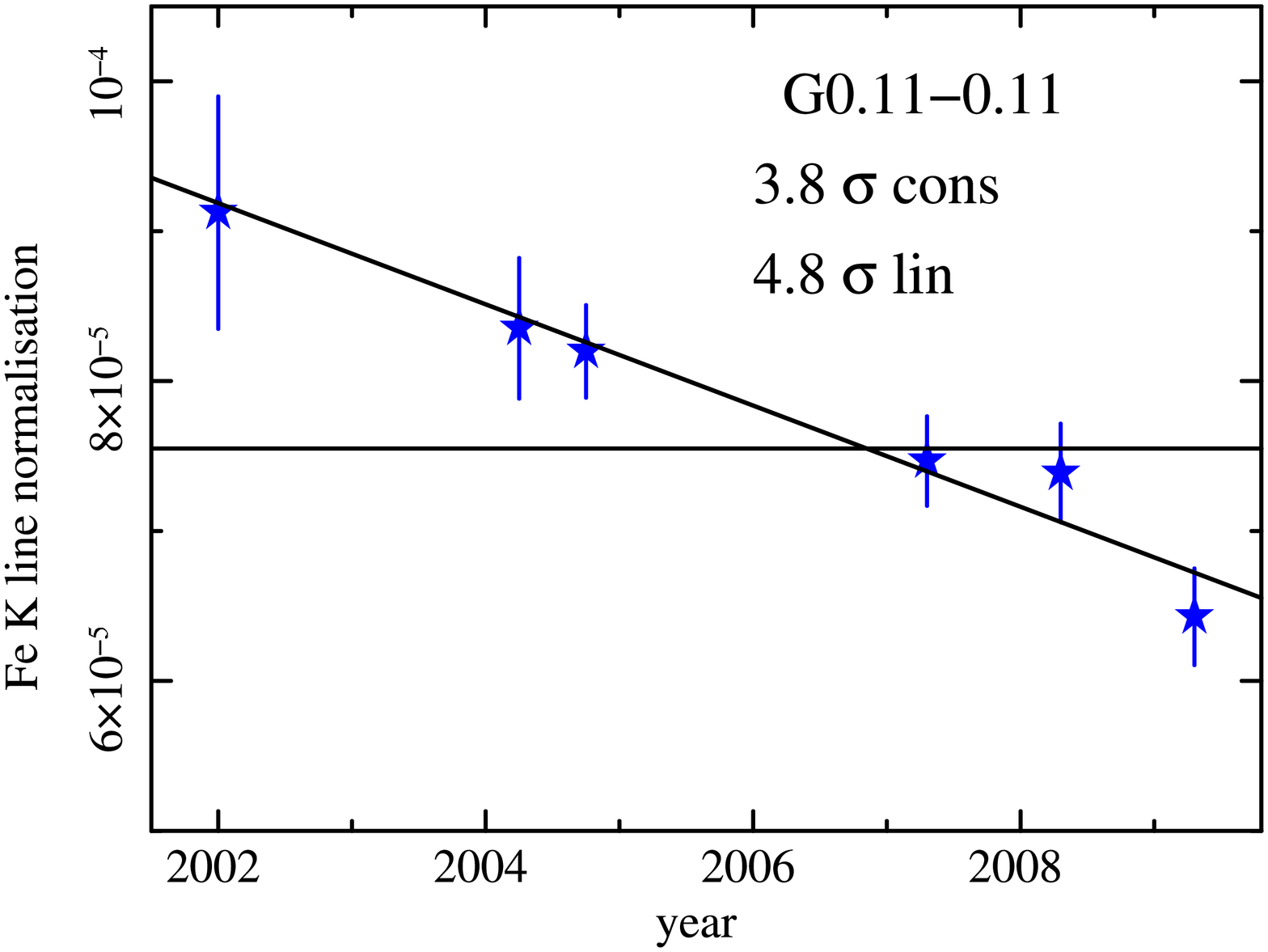}
\includegraphics[scale = 0.26]{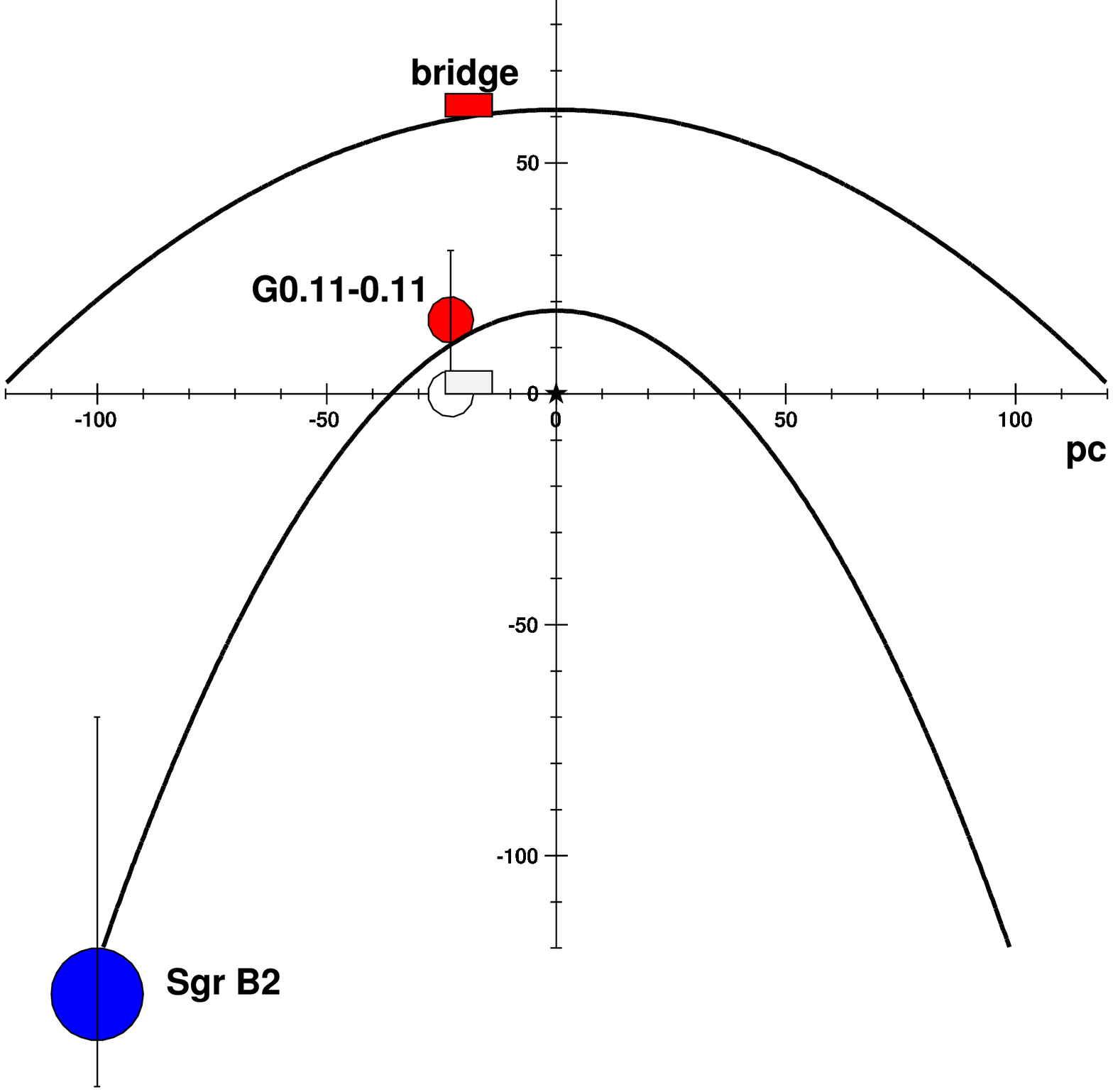}
\includegraphics[scale = 0.67]{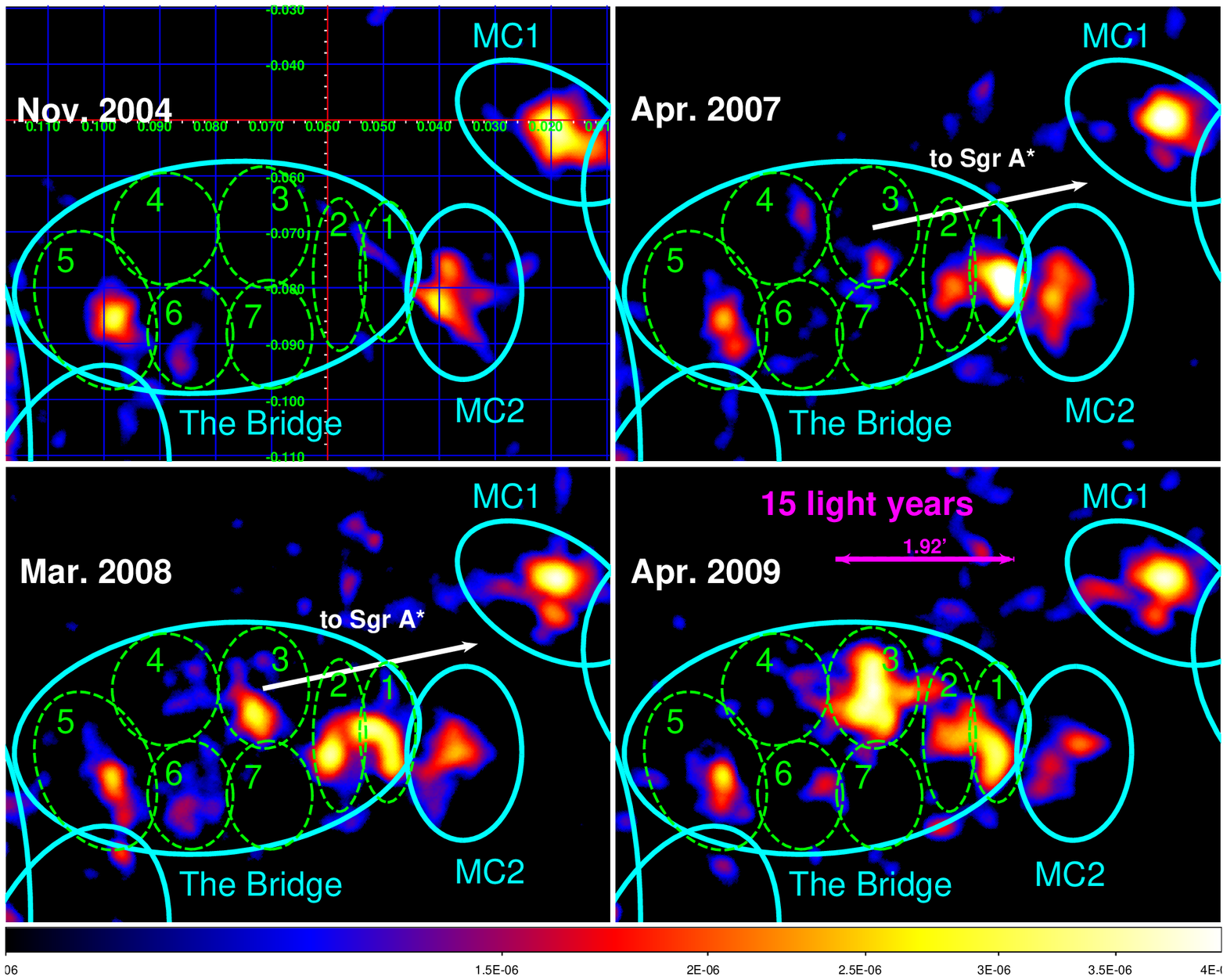}
\caption{{\it Left panel}: Fe K intensity light curve of G0.11-0.11. The reflection 
features in this MC are variable (at a significance level of about 3.8 
$\sigma$) with a linear decay. 
{\it Right panel}: Sketch of the face--on view of the Galactic 
Plane as seen from the direction of the Galaxy pole. Sgr A* is indicated by a star. 
Galactic East is toward negative abscissa and the direction toward the Earth 
is bottom, at negative ordinate. A Sgr A* light front observed from the Earth 
appears as a parabola. Surprisingly the same parabola hits Sgr B2 and 
G0.11-0.11, in agreement with reflecting the same Sgr A* flare (about 100 
years ago).
{\it Lower panel}: Fe K$\alpha$ continuum subtracted mosaic images of 
the bridge region. A brightening of 
the bridge 1, 2, 3 and 4 is clear. Such variation occurs in a time-scale of 
about 2-4 years, but in a region of about 15 light years. 
This apparent super-luminal motion can be explained if the bridge 
is illuminated by a bright (L$>1.3\times10^{38}$ erg \s) and distant 
($>$15 pc) X-ray source active for several years. The primary source seems 
to be in the direction of Sgr A*.}
\label{lc_G011}
\end{figure}

\section{Spatial variations}

Fast and important variations occur also in the MC spatially 
located between G0.11-0.11 and feature 2 (see Muno et al. 2007) and called "the 
bridge" (Armstrong et al. 1985; Sakano et al. 2006; Ponti et al. 2010). 
The different parts of the bridge (from bridge 1 to 4, see Fig. 1) 
show similar Fe K intensity variations (with significance between 8-12 $\sigma$,
each). Moreover the light curve evolutions are consistent with being the 
same, with only a time delay between them. 

\subsection{Discovery of superluminal echo in the bridge region}

Fig. 1 shows the continuum subtracted, exposure corrected 
EPIC (pn+MOS) image of the bridge in the Fe K band (6.28-6.53 keV). 
The continuum is measured in the 4.5-6.28 keV band and subtracted after 
the extrapolation in the Fe K band assuming a power law spectral index 
emission of $\Gamma=2$. 

Significant Fe K intensity variations occur in the regions bridge 1, 2 and 3.
In 2007 the bridge 1 lights up, becoming as intense as the region MC2.
This variation evolves toward North East in 2008. While, in 2009, the bridge 
3 region has the higher intensity.
This behaviour suggests a connection/evolution of the rising up of the 
different regions, suggesting an origin tied to the propagation of an event 
in the bridge. 
Nevertheless, the emitting regions are causally disconnected. 
The variations, in fact happens in 2-4 years, while they are separated by at 
least 15 light years (being the bridge located at the GC).

This excludes an internal source, or low energy cosmic rays as possible 
candidates for the observed variation.
The different regions have similar intensity variations, this implies that the 
source has to be more distant that several times the bridge length, otherwise 
a detectable modulation (due to the dependence of the intensity with the 
square of the distance) should be detected. Thus, if the source is more distant 
than 15 pc, it has to be brighter than L$>1.3\times10^{38}$ erg \s. This value 
is close to the Eddington luminosity for a stellar mass black hole. 
We also note that the illumination starts from the Galactic west of the 
bridge and propagates towards the Galactic east, diffusing in 
the northern part of the bridge. This strongly suggests that the 
illuminating source is located east of the bridge, slightly north of it. 
This is the direction toward Sgr A*. 

Assuming that the bridge is reflecting a Sgr A* flare with the same luminosity
illuminating Sgr B2, then the bridge has to be located about 60 pc behind 
Sgr A*. Thus, in this hypothesis, the period of activity of Sgr A* has to have lasted 
at least a few hundred years.

\acknowledgements 
The work reported here is based on observations obtained with
XMM-Newton, an ESA science mission with instruments and contributions
directly funded by ESA Member States and NASA. GP thanks ANR for 
support (ANR-06-JCJC-0047). The authors thank the team of astronomers 
that helped in building such a long monitoring campaign of the GC region.

\end{document}